\documentclass{elsart1p}
\usepackage{epsfig}
\usepackage{ulem,color,graphics}
\usepackage{amssymb}
\usepackage{graphicx}
\usepackage{epsfig}
\usepackage{bm}

\def\gsim{\:\raisebox{-0.5ex}{$\stackrel{\textstyle>}{\sim}$}\:}
\def\etal{{\it et al.}}
\def\go{\rightarrow  }
\def\be{\begin{equation}}
\def\ee{\end{equation}}
\def\br{\begin{eqnarray}}
\def\er{\end{eqnarray}}
\def\brn{\begin{eqnarray*}}
\def\ern{\end{eqnarray*}}
\def\rf#1{{(\ref{#1})}}

\def\bit{\begin{itemize}}
\def\eit{\end{itemize}}

\def\gsim{\:\raisebox{-0.5ex}{$\stackrel{\textstyle>}{\sim}$}\:}
\def\ie{{\it i.e., }}

\def\rf#1{{(\ref{#1})}}

\def\go{\rightarrow  }

\newcommand{\Mass}{\mathrm{M}}

\def\rf#1{{(\ref{#1})}}

\def\be{\begin{equation}}
\def\ee{\end{equation}}
\def\br{\begin{eqnarray}}
\def\er{\end{eqnarray}}

\newcommand{\He}{$^5_\Lambda$He}
\newcommand{\he}{$^4_\Lambda$He}

\begin{document}

\begin{frontmatter}
\title{Nonmesonic weak  decay spectra  of light hypernuclei }
\author{Franjo Krmpoti\'c}
\address{Instituto de F\'isica La Plata, CONICET, 1900 La Plata,
Argentina, and Facultad de Ciencias Astron\'omicas y Geof\'isicas,
Universidad Nacional de La Plata, 1900 La Plata, Argentina,\\
and\\
Instituto de F\'isica Te\'orica, Universidade Estadual Paulista\\
Rua Dr. Bento Teobaldo Ferraz 271-Bloco II,
S\~ao Paulo, SP 01140-070, Brazil.}

\begin{abstract}
The nonmesonic weak decay spectra of light hypernuclei have been evaluated in a
systematic way. As  theoretical framework we adopt the independent particle
shell model with three different one-meson-exchange transition potentials.
Good agreement with  data is obtained for proton and neutron kinetic energy  spectra
of $^4_\Lambda$He, and $^5_\Lambda$He, when the recoil effect is considered. The coincidence
 spectra of  proton-neutron pairs are also accounted for quite reasonably,
but  it was not possible to reproduce the data for  the neutron-neutron pair spectra.
It is  suggested that the $\pi+K$ meson-exchange model with soft monopole form factors
could be a good starting point for describing the dynamics responsible for the  decays
of these two hypernuclei. The $^4_\Lambda$H~ spectra are also presented.
 \end{abstract}

\end{frontmatter}
\section{Introduction}
Investigations of
unusual nuclear properties, such as nontrivial values of  flavor quantum
numbers (strangeness, charm or beauty), or large isospin (so
called neutron rich isotopes) are of continuous  interest.
A particularly interesting phenomenon in nuclear physics is the
existence of nuclei containing strange baryons. The lightest
hyperons are stable against strong and electromagnetic decays, and
as they do not suffer from Pauli blocking by other nucleons they
can live long enough in the nuclear environment  to become bound.
When a
hyperon, specifically a $\Lambda$-hyperon with strangeness ${\sf S}=-1$, replaces one of the
nucleons in the nucleus, the composed system acquires different properties
from that of the original one, and is referred as hypernucleus.

One such  very important new  property is
 the additional binding.
For instance, while the
one-neutron separation energy in $^{20}$C is $1.01$ MeV, it is
$1.63$ MeV in $^{21}_\Lambda$C, and $^{6}_\Lambda$He is bound
while $^{5}$He is unbound.
 As a consequence the neutron drip line is modified, and  the extended  three-dimensional
  $({\sf N,Z,S})$ domain of  radioactivity becomes
 even richer in elements than the
ordinary neutron-proton domain $({\sf N,Z})$.
Because of this glue attribute of hypernuclei
the $\Lambda N$ interaction is  closely related to the inquiry on
the  existence of strange quark matter and its fragments, and
strange stars (analogues of neutron stars), which  makes  the
hypernuclear physics  also relevant for astrophysics and
cosmology.

 Another remarkable property of $\Lambda$-hypernuclei is the occurrence of
  the nonmesonic weak decay   (NMWD): $\Lambda
N\go nN$ with $N=p,n$, which  is the main decay channel for medium and heavy hypernuclei.
 This decay  takes place only within nuclear  environment,
and is  the unique opportunity that nature offers us to inquire about
strangeness-flipping   interaction between baryons.

The knowledge  of strange hadrons carrying an additional flavour
degree-of-freedom is essential for understanding the low-energy
regime of the quantum chromodynamics (QCD), which is the field theory of strong
interactions among quarks and gluons, and  is widely
used  at very high energies. Yet,  at  energy scale of the nucleon mass
the hadrons represent complex many body systems, making difficult  the
QCD description due to  the non-perturbative nature of the theory.

Ergo the NMWD dynamics is frequently handled by the one-meson-exchange (OME) models
~\cite{Du96,Pa97,It02,Ba02,Kr03,It03,Ba03,Ba07,It08}, among which
  the exchange of  full pseudoscalar ($\pi, K, \eta$) and vector
($\rho,\omega,K^*$) meson octets (PSVE) is  the most used one.  This model is based on the original idea of
 Yukawa that the $NN$  interaction at long distance is due to the one-pion-exchange (OPE),
while the weak coupling constants are obtained from soft meson theorems and
$SU(6)_W$~\cite{Du96,Pa97}.
The dominant role is being played  by the
exchange of  pion and kaon mesons (PKE). We have recently shown that the $\pi+K$ meson exchange
potential with soft dipole form factors (SPKE) reproduces fairly well
both $s$-shell ~\cite{Ba09}, and  $p$-shell ~\cite{Kr12} NMWD.

A  hybrid mechanism has been also meticulously  used by the Tokyo group~\cite{In97,Ok98,Sa00,Sa02,Ok05,Sa05}
to describe the NMWD in $A = 4$, and $5$ hypernuclei.
This group represented the short
range part of the $\Lambda N$ weak interaction  by the direct quark
(DQ) weak transition potential, while the longer range interactions are assumed to come from
the exchange of $\pi+K$ mesons with soft pion form factor
\footnote{McKeller and
Gibson employed a very soft $\pi$ of $0.63$ GeV~\cite{Mc84}.}.

The $\Lambda$-hypernuclei are mainly produced by the $(K^-, p^-)$
and the $(p^+,K^+)$ strong reactions,  and disintegrate by the
weak decay with the rate
\[
\Gamma_W = \Gamma_M +\Gamma_{NM},
\]
where $\Gamma_M$ is  decay  rate for the mesonic (M) decay $\Lambda \rightarrow \pi N$,  and  $\Gamma_{NM}$
is the rate for the nonmesonic (NM) decay, which can be induced either by one bound nucleon ($1N$),
$\Gamma_1(\Lambda N  \rightarrow nN)$,
or by two bound nucleons ($2N$), $\Gamma_2(\Lambda NN \rightarrow nNN)$,
\ie
\[
\Gamma_{NM}=\Gamma_1+\Gamma_2,
\]
with
\[
\Gamma_{1}=\Gamma_p+\Gamma_n,
\hspace{.4cm}\Gamma_{2}=\Gamma_{nn}+\Gamma_{np}+\Gamma_{pp}.
\]
In the evaluation of $\Gamma_{NM}$, and the corresponding spectra,  are used the Shell Model (SM)
and the Fermi Gas Model (FGM) for $\Gamma_1$ (within
the two-particle phase space)~\cite{Du96,Pa97,It02,Ba02,Kr03,It03,Ba03,Ba07,It08,Ba09,Kr12,In97,Ok98,Sa00,
Sa02,Ok05,Sa05,Mc84,Bau09a,Bau09b,Bau09c,Ba10a}, while for  $\Gamma_2$ (within
the three-particle phase space) only the later model has been employed so far
\cite{Bau09a,Bau09b,Bau09c,Ba10a}.

In our previous work~\cite{Ba09} we have analyzed the Brookhaven
National Laboratory (BNL) experiment E788 on
$^4_\Lambda$He~\cite{Pa07}, involving: 1)
the  single-proton spectra $S_{p}(E)$,  and
total neutron spectra $S_{nt}(E)=2S_{n}(E)+S_{p}(E)$,  as a
function of corresponding one-nucleon kinetic energies $E_N$, and 2)
two-particle-coincidence spectra, as  a function of: i) the sum of
kinetic energies $E_n+E_N\equiv E_{nN}$, $S_{nN}(E)$, and ii)
the opening angle $\theta_{nN}$, $S_{nN}(\cos\theta)$.
This has been done within  a simple theoretical framework,
based on the independent-particle SM (IPSM) for the
$1N$-induced NM weak decay spectra. That is, we have disregarded both
the $2N$-NM decay, and the final state interactions (FSIs),
which  is consistent with the upper limits
 $\Gamma_{2}/\Gamma_{W}\le 0.097$, and $\Gamma_{NM}^{FSI}/\Gamma_{W}\le 0.11$
 established in ~\cite{Pa07} with a $95\%$  CL.
\footnote{We note that  experimentally it is very difficult to distinguish
between these two effects and very likely  $(\Gamma_{2}+\Gamma_{NM}^{FSI})/\Gamma_{W}
\sim 10\%$.}
Moreover, the calculated spectra were normalized to the
experimental ones. For instance, in the case of single proton
spectrum $S_{p}(E)$,  the number of protons  $\Delta {\rm N}_p^{exp}(E_i)$  measured  at energy $ E_i$
within a fixed proton energy bin $\Delta E$,
is confronted with the calculated number
 \be
\Delta {\rm N}_p(E)= {\rm N}_p^{exp}\frac{S_p(E)}{\Gamma_p}\Delta E,\hspace {0.5cm}{\rm N}_p^{exp}=\sum_{i=1}^{m} \Delta {\rm N}_p^{exp}(E_i),
\label{1}\ee
where
${\rm N}_p^{exp}$
is the total number of measured protons, while
 $S_p(E)$ and $\Gamma_p=\int S_p(E)dE$ are   evaluated theoretically.
\footnote{Corrections for detection acceptance and threshold are discussed in ~\cite{Ba09}. }
Note that: a) while   the   $\Delta {\rm N}_p^{exp}(E_i)$ are defined only at $m$ experimental energies $E_i$,
the quantity $\Delta {\rm N}_p(E)$ is a continuous function of $E$, and b)
the condition ${\rm N}_p={\rm N}_p^{exp}$ is automatically fulfilled  when $\Gamma_p=\Gamma_p^{exp}$.

To describe the NMWD dynamics three different OME potentials have been tested in Ref.
~\cite{Ba09}, namely:

P1)  The full PSVE potential that includes
  the exchanges of  nonstrange-mesons $\pi,\rho,\omega$, and $\eta$, and strange-mesons $K$, and $K^*$
with the weak coupling
constants from the Refs.
~\cite{Du96,Pa97,Ba02},

P2) The PKE model, with usual
cutoffs $\Lambda_\pi= 1.3$ GeV, and  $\Lambda_K=1.2$ GeV, and

P3) The SPKE potential, which corresponds to the
 cutoffs $\Lambda_\pi= 0.7$ GeV and  $\Lambda_K=0.9$ GeV.

The calculated spectra shown in Figs. 2, 3, and 4 in Ref.~\cite{Ba09} correspond to  the parametrization P3,
but almost identical spectra are obtained with the remaining two parametrizations, which
  is due to
the fact that the shapes of all  spectra are  basically  tailored by the kinematics, depending very weakly on the dynamics,
which cancels out almost wholly  in \rf{1} when  $S_p(E)$ is divided by $\Gamma_p$.
Moreover, the fairly good agreement between the data and calculations indicates that the
framework employed (IPSM plus the two-particle phase space and the recoil effect) could be appropriate for describing
the kinematics in  the NMWD of \he.
It  indicates as well  that, neither  the FSI, nor the two-nucleon induced
decay processes play a very significant role.
Needless to stress that we   study the NMWD  in order  to understand
 its mechanism,  and to  disentangle in this way  the strangeness-flipping interaction among baryons, while
   the FSIs, together  the $2N$-NM decay,  are to some extent just  undesirable and inevitable complications.

To  test not only the kinematics but also the dynamics the comparison between the experimental and calculated
spectra has to be done differently, as already pointed out in Refs. ~\cite{Kr12,Go11,Go11a}.
This is done by confronting
 straightforwardly the corresponding
spectra. For instance, for the  case considered above one compares the calculated
spectrum $S_p(E)$ with the experimental one defined as:
\br
S_p^{exp}(E_i)&=&\left.\frac{\Gamma_{W}}{{\rm N}_{W}}
 \frac{\Delta{\rm N}_p(E_i)}{\Delta E}\right|_{BNL,FINUDA}
=\left.\frac{\Gamma_{NM}}{{\rm N}_{NM}}
 \frac{\Delta{\rm N}_p(E_i)}{\Delta E}\right|_{KEK}.
 \label{2}\er

The difference between the first and second terms in the right side is due to
different  normalizations  of $\Delta{\rm N}(E_i)$. The BNL
and FINUDA groups do it in relation with the total number of weak
decays ${\rm N}_{W}$, while the KEK data are normalized to the
number of NM decays ${\rm N}_{NM}$. In both cases the experimental proton
transition rate, derived from the proton kinetic energy spectrum $S_p(E)$, is defined as
\br
\Gamma_p^{exp}={\Delta E}\sum_iS_p^{exp}(E_i),
 \label{3}\er
and similarly those obtained from $S_{nt}(E)$, $S_{nN}(\cos\theta)$,
$S_{nN}(E)$, and $S_{nN}(P)$. The neutron decay rate derived from
the total neutron kinetic energy spectrum is defined as:
$\Gamma_n^{exp}=(\Gamma_{nt}^{exp}-\Gamma_p^{exp})/2$.
For $\Gamma_{W}$
we use the relationship
 \[
{\Gamma}_{W}(A)=(0.990\pm 0.094)+(0.018\pm0.010)~A,
\]
which was determined in Ref.~\cite{Ag09} for all measured hypernuclei in the mass range $A=4-12$.

We will analyze here
as well the spectra $S_{nN}(P)$ as a function of the center of mass (c.m.) momentum
$P\equiv P_{nN}=|\textbf{P}_{nN}|=|\textbf{p}_n+\textbf{p}_N|$.
More, we
will  seize the opportunity to discuss
 the spectra of other two
light hypernuclei, namely  of $^4_\Lambda$H, and $^5_\Lambda$He,
taking care of  measurements done on the latter at
KEK~\cite{Ok04,Ka06,Ki06},
  and FINUDA~\cite{Ag10}.
Theoretical expressions for different NM weak decay spectra
within the IPSM that are used here have been
presented  previously~\cite{Ba09,Ba08,Kr10,Kr10a,Go11,Go11a}
 and will not be repeated here.

\section{Results}
\begin{figure}[htpb]
\vspace{1.5cm}
\begin{tabular}{cc}
\includegraphics[width=4.7cm,height=4.5cm]{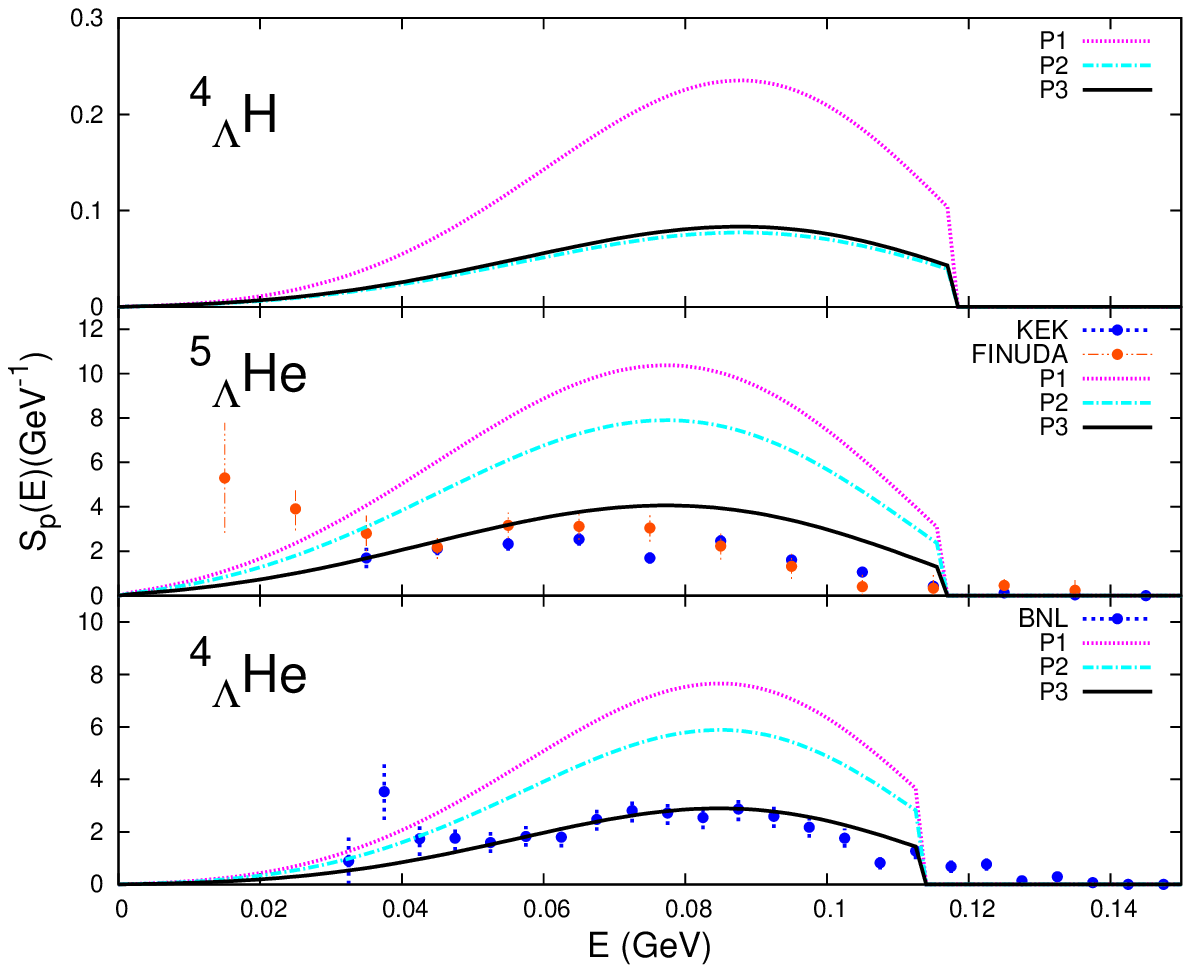}&
\hspace{2cm}\includegraphics[width=4.7cm,height=4.5cm]{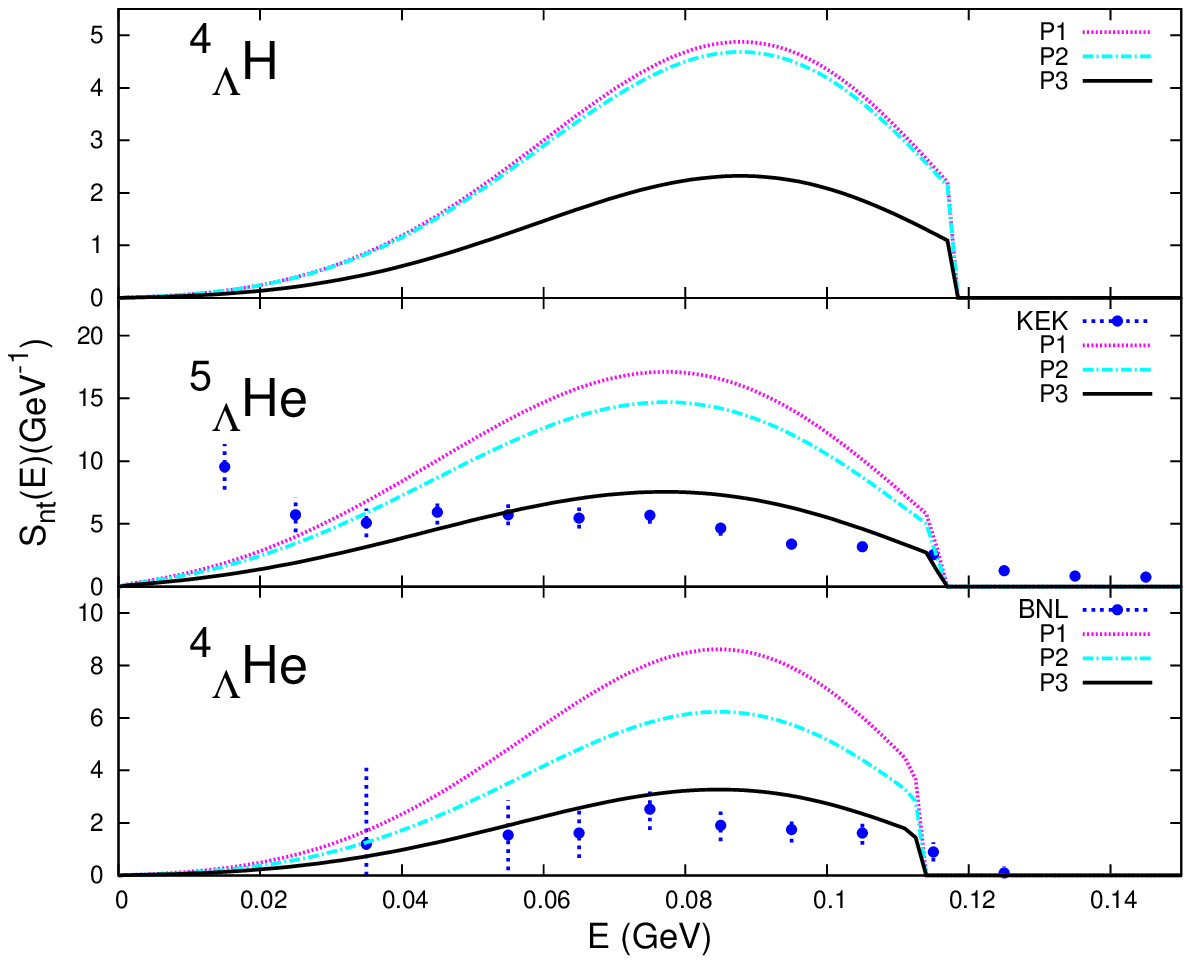}
\end{tabular}
\vspace{0.51cm}
 \caption{\label{F1}(Color online)
Calculations  of  the proton kinetic energy spectrum $S_p(E)$ (left panel),
and the total neutron spectrum $S_{nt}(E)=S_p(E)+2S_n(E)$ (right panel), for three different
parametrizations are confronted with  measurements done at BNL
 for \he~\cite{Pa07}, and at  KEK~\cite{Ok04}, and
FINUDA~\cite{Ag10}  for \He.}
\end{figure}

To describe the NMWD dynamics, we will employ  the OME potentials
that are  listed above.
The short range correlations (SRC) acting on
 final  $nN$ states are incorporated a posteriori
 phenomenologically  through Jastrow-like SRC functions, as used
  within both finite nuclei~\cite{Pa97,Ba02,Kr03,Ba03,Ba07},
 and FGM calculations
 ~\cite{Ba04,Bau09a,Bau09b,Bau09c,Ba10a,Bau11}.
The  size parameter $b$, which is the most important nuclear structure parameter~\cite{Kr10a},
was estimate to be $b=\frac{1}{2}\sqrt{\frac{2}{3}}\left(R_N+R_\Lambda\right)$ where $R_N$ and
$R_\Lambda$ are, respectively, the root-mean-square distances of the the nucleons and the
$\Lambda$ from the center of mass of the hypernucleus. This yields $b(^4_\Lambda$H)$=1.57$ fm,
$b(^4_\Lambda$He)$=1.53$ fm, and $b(^5_\Lambda$He)$=1.33$ fm
~\cite{Ko06}.

In Fig. \ref{F1} are shown the theoretical results for: a)  the proton kinetic energy $S_p(E)$
developed by the decay $\Lambda p\go np$ (left panel), and b) the total neutron spectrum
$S_{nt}(E)$,  induced
 both by protons, $\Lambda p\go np$, and by neutron, $\Lambda n\go nn$, yielding, respectively,
 the spectra $S_p(E)$, and $S_n(E)$.
The BNL data for \he~\cite{Pa07}, and the KEK~\cite{Ok04}, and
FINUDA~\cite{Ag10} data for \He~ are shown as well. Within the parametrizations P1 and P2 the theory
strongly overestimates all the data.
In turn the SPKE potential  reproduces both  \he~ spectra quite well, but not so well in the case of \He.
Here the spectra are undervalued  for energies $E\gsim 0.04$ GeV, while they are somewhat overvalued
 for higher energies.
 This relatively minor discrepancy could  be attributed to lack of FSIs in
 the theory whose main effect is to shift a portion of the  transition strength
 $\Gamma_1$ from high energies towards  low energies. The inclusion in the calculations
 of  $\Gamma_2$, which contributes dominantly at low energies, could also improve  the agreement.

As expected,  the theoretical spectra $S_{p}(E)$,
are peaked around the half of the corresponding liberated energy ($Q$-value):
 $\Delta_p=\Delta  +
\varepsilon_{\Lambda} + \varepsilon_{N}$, where
 $\Delta=\Mass_\Lambda-\Mass=0.1776$ GeV is the
 $\Lambda-N$ mass difference, and  $\varepsilon$'s are the
single-particle energies. However,  in light hypernuclei
the  spectrum shape is  not exactly  that of a symmetric inverted bell,
since the  single-proton kinetic energy reaches rather abruptly its maximum value
\footnote{There is an
mistake in Eq. (13) of Ref.~\cite{Ba09}.}
 \be
E^{\tiny max}_{p}=\frac{A-2}{A-1}\Delta_p,
 \label{4}\ee
with $A$ being the nuclear mass number, which is significantly smaller than  the corresponding $Q$-value,
as a consequence of  energy conservation, and the recoil effect. One gets,
 in units of GeV:
\[ \begin{array}{ccr}
\mbox{$\Lambda$-nucleus} &\Delta_p&E^{\tiny max}_{p} \\
\mbox{$^4_\Lambda$H} & 0.175&0.117 \\
\mbox{\he} &0.165 &0.110\\
\mbox{\He} & 0.154 & 0.115\end{array}
\]
In the absence of the recoil effect  $E^{\tiny max}_{p}\equiv\Delta_p$, and therefore it can
be concluded that this effect is very important in  defining the shape of the light hypernuclei spectra.
 Similar results are obtained for the  neutron spectrum $S_n(E)$, as well as
for the total neutron spectrum $S_{nt}(E)$.

It is worthy of note that, when the SPKE potential is used, the predicted
$S_{nt}(E)$ spectrum in $^4_\Lambda$H~ turns out to be of the same order of magnitude
as those in  \he, and \He. Therefore it would be very interesting to measure the $^4_\Lambda$H~neutron spectrum
to verify whether this prediction is fulfilled.

\begin{figure}[htpb]
\vspace{2.5cm}
\begin{tabular}{cc}
\includegraphics[width=4.7cm,height=4.5cm]{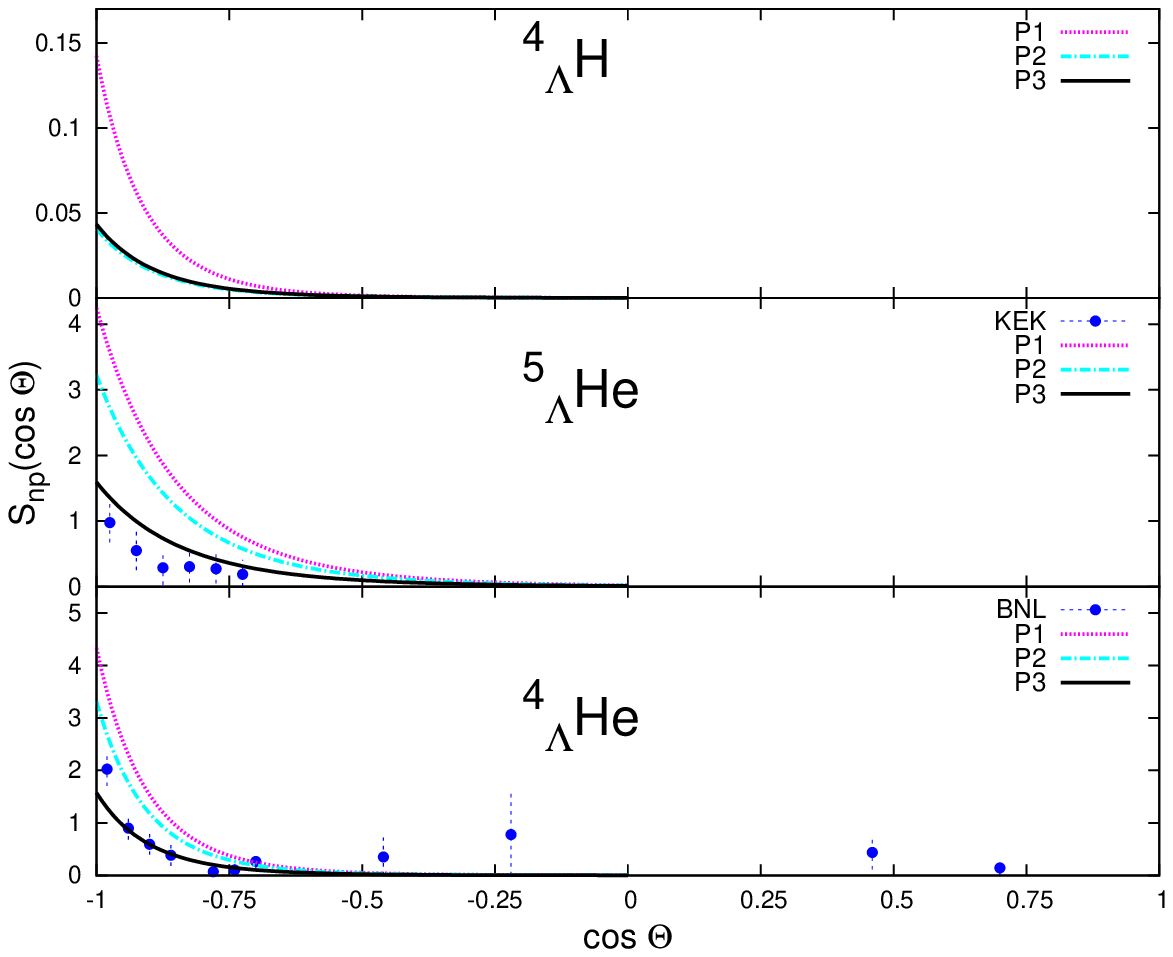}&
\hspace{2cm}\includegraphics[width=4.7cm,height=4.5cm]{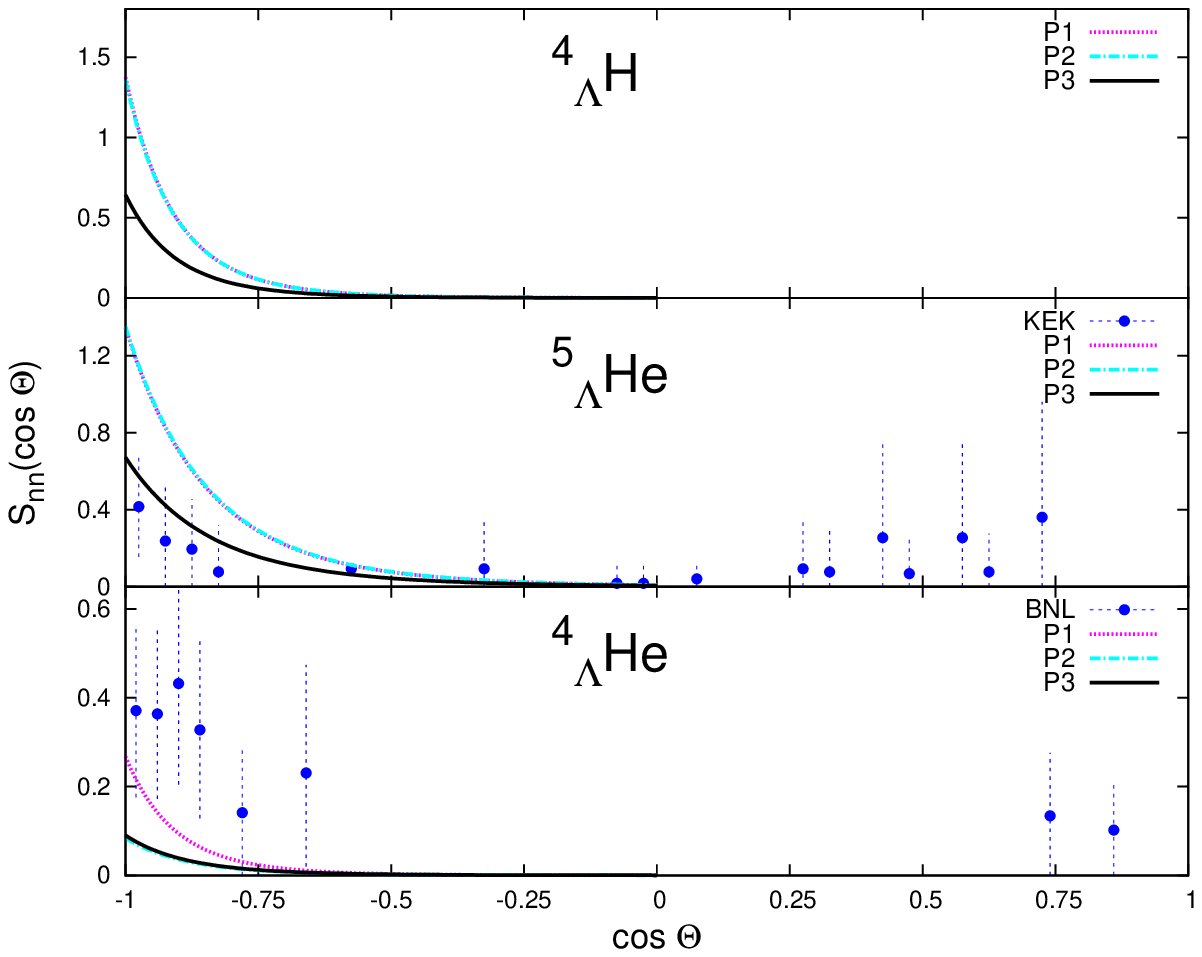}
\end{tabular}
\vspace{0.51cm}
 \caption{\label{F2}(Color online)
Calculations  of
the opening angle
correlations  of neutron-proton pairs $S_{np}(\cos\theta)$
(left panel), and of neutron-neutron pairs $S_{nn}(\cos\theta)$
(right panel),
 for three different
parametrizations are confronted with  measurements done at BNL
 for \he~\cite{Pa07}, and at  KEK~\cite{Ka06,Ki06} for \He.}
\end{figure}
In Fig.~\ref{F2} are  exhibited the calculations of  opening angle distribution for
the $np$ (left panel) and $nn$ (right panel) pairs, which, as expected, emerge roughly
back-to-back with a separation angle near $\theta_{np}=180^\circ$.
 First,  it should be noticed
  that the approaches P1 and P2  yield virtually
 identical results for $S_{nn}(\cos\theta)$  in $^4_\Lambda$H~ and \He,  to the point that they can  not be distinguished visually.
Something very similar happens
 with  $S_{np}(\cos\theta)$ in $^4_\Lambda$H, and   $S_{nn}(\cos\theta)$ in \he~
 within the parametrizations P2 and P3.
The interpretation of the smilarity  between the P1 and P2 results is that the contributions of heavy mesons have little effect in this case.
Similarly, one can say that when P2 and P3 models produce similar results the size of the  form factors is of little importance.
All these  overlays will be repeated in other spectra in coincidence.

 Again the best agreement between  the theory, and: i)  the  BNL
 spectrum  $S_{np}(\cos\theta)$ in \he~\cite{Pa07}, and  ii)  both  KEK spectra $S_{nN}(\cos\theta)$   in \He~\cite{Ka06,Ki06}
 is achieved with the SPKE potential,
which  yet overestimates slightly  the latter data.
On the other hand, the theory is not able to account for  the $nn$ coincidence spectra
  measured in the BNL experiment
 on \he~\cite{Pa07}, nor for the KEK data~\cite{Ka06,Ki06} in \He~ at angles
  $\theta_{nn}<90^\circ$. We do not believe that the FSIs and/or the 2N-NM decay are capable to solve these problems,
since these effects were not able to correctly describe the $S_{nn}(\cos\theta)$ spectra in $^{12}_\Lambda$C
(see Fig. 4 in Ref. ~\cite{Bau11}, and  Fig. 7 in Ref. ~\cite{Go11a}).
 At this point it might be interesting to compare  the concordance obtained
in Fig. 3 of Ref.~\cite{Ba09} for the $nn$ angular coincidences in \he~ with the strong
 disagreement for the same observable exhibited  in Fig.~\ref{F2}. The answer to this apparent contradiction can be found
in  the fact that  in ~\cite{Ba09} has been used the relationship \rf{1} where all information on
the dynamics is washed out almost entirely.

It is also likely that in the case of \he~ the kinematics of the final state does not reflect the
the real situation. In fact, one should also keep in mind that  in the decay
channel $^4$He$\go (2p)+n+n$ the remaing two protons are not in a bound state, and that one is faced with the
so-called four-body problem~\cite{Ju01}. It is much harder to find a plausible explanation for
the discrepancy in  \He, which may be indicative of $nn$ coincidences originated from sources other
than  $\Lambda n $ decays, as already suggested in Ref.~\cite{Pa07}.
\footnote{The decay of \he~ and \He~  via two-body channels \he$\go d+d, p+t$, and \He$\go d+ t$, have been  measured
recently by FINUDA~\cite{Ag09a} to be of a few percent of the one-proton induced
decay. Nothing is said in this work about an unbound four-nucleon final state.}

\begin{figure}[htpb]
\vspace{2.5cm}
\begin{tabular}{cc}
\includegraphics[width=4.9cm,height=4.5cm]{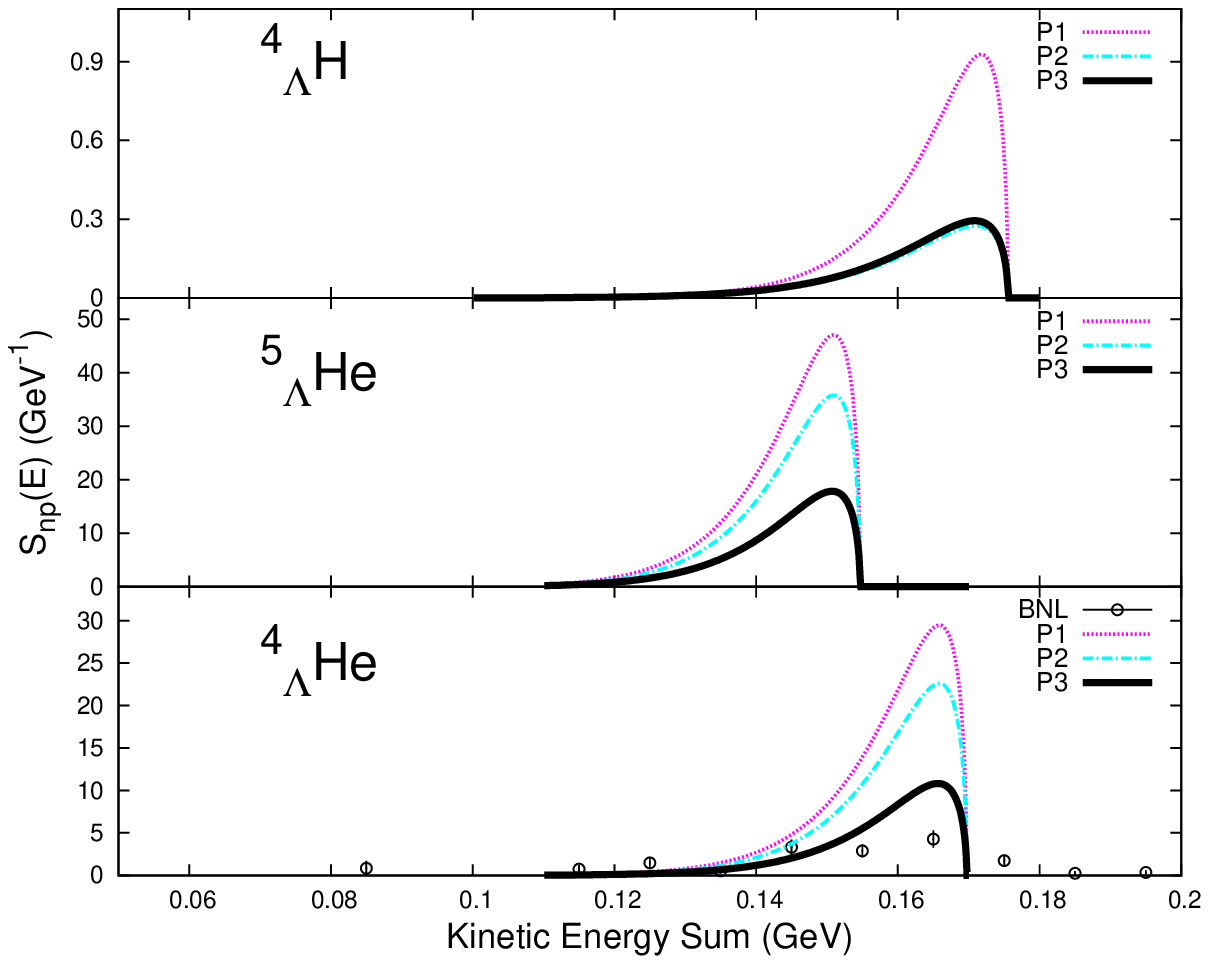}&
\hspace{2cm}\includegraphics[width=4.9cm,height=4.5cm]{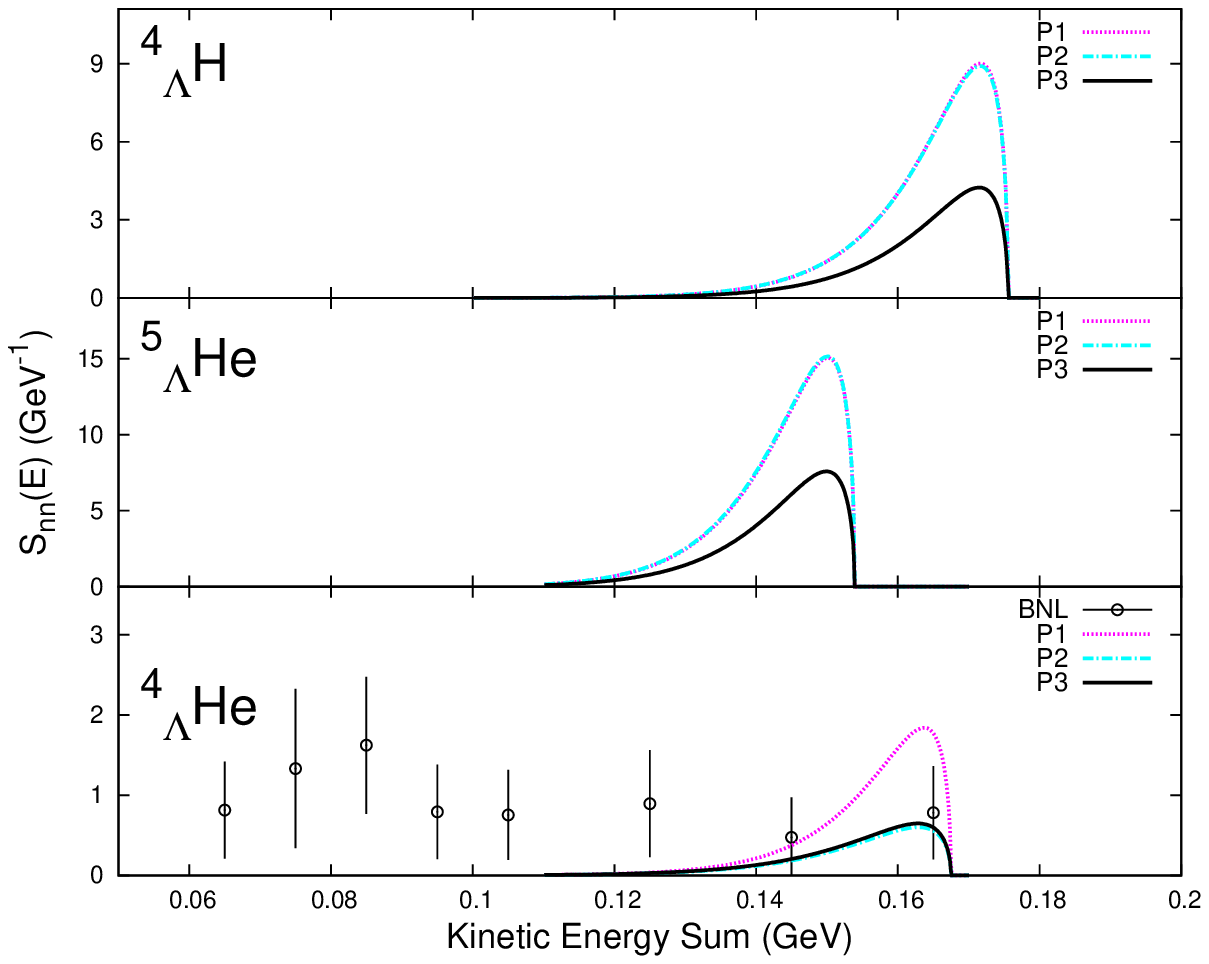}
\end{tabular}
\vspace{0.51cm}
 \caption{\label{F3}(Color online)
Calculations  of   the kinetic energy sum
correlations  $S_{nN}(E)$ of $np$ (left panel) and $nn$ (right panel) pairs   for three different
parametrizations are confronted with the measurements done so far at BNL
 for \he~\cite{Pa07}.}
\end{figure}

Next, we discuss the spectra for the kinetic energy sums $E\equiv E_{nN}=E_n+E_N$   in
NMWD  $\Lambda N\go nN$. As demonstrated by Barbero \etal~\cite{Ba08}
within the IPSM these spectra contain one or more
peaks, the number of which is equal to the number of shell-model orbitals
$ 1s_{1/2}, 1p_{3/2}, 1p_{1/2}, \cdots$
that are either fully or partly occupied.
Before including the recoil, all these peaks would be just spikes
at the corresponding Q-values $\Delta_N$. With the recoil effect
they develop rather narrow widths $\sim [b^2M(A-2)]^{-1}$.
For the s-shell
hypernuclei there is  only one such peak, and the spectra behave in this case as
\br
S_{nN}(E \cong\Delta_N)&\sim& \sqrt{(\Delta_N-E)(E-\Delta'_N)}
e^{-M(A-2)(\Delta_N-E)b^2},
\label{5}
\er
where
\be
\Delta'_N=\Delta_N\frac{A-2}{A},
\label{6}\ee
is the minimum kinetic energy of the emitted pair nN. That is, the spectra  $S_{nN}(E)$ are
restricted within the energy intervals $\Delta'_N<E<\Delta_N$.

In the left and right panels of Fig. \ref{F3}  are displayed, respectively, the calculated correlated  spectra
as a function  of  sums  of kinetic energies
of $np$ pairs, $S_{np}(E)$, and  $nn$ pairs, $S_{nn}(E)$.
The measurements done so far at BNL
 for \he~\cite{Pa07} are also exhibited.
 The experimental $S_{np}(E)$ spectrum of \he~ agrees fairly well
 with the theory when the parametrization P3 is used.
Moreover, as seen from the right panel of Fig. \ref{F3}, the data  of the  \he~
 spectrum  $S_{nn}(E)$ differ strongly from the theoretical calculation,
 which is consistent with the similar deviation of the spectrum $S_{nn}(\cos\theta)$
in Fig. \ref{F2}. Again, the reason for this discrepancy may be nn coincidences that do not come from NMWD.
In this regard, one should remember  that is valid the relationship
\br
\Gamma_n&=&\negthinspace\int S_{nn}(\cos\theta)d\cos\theta
=\int S_{nn}(E)dE,
\label{7}\er
as well as,  that the experimental $\Gamma_n$ for  \he~ is very small ($\Gamma_n/\Gamma_W\leq 0.018$~\cite{Pa07}).

One has to mention as well that the KEK group has measured
 the $nN$ distributions of the sums of kinetic energies  in \He~\cite{Ka06,Bh07}.
But we can not
compare quantitatively their results with our calculations, since in their  work the number of measured pairs
are not normalized to the number of NM decays $ {{\rm N}_{NM}}$. Nevertheless,  we can say
that, while theoretically both $S_{nN}(E)$ spectra in \He~ are peaked up at $\sim 0.15$ GeV
with a width
of $\cong 20$ MeV, experimentally is observed a sharp peak at $\sim 0.14$ GeV in the $np$
spectrum, while  a wide bump within the energy range $0.08-0.15$ GeV has been
detected in the $nn$ spectrum.

\begin{figure}[htpb]
\vspace{2.5cm}
\begin{tabular}{cc}
\includegraphics[width=4.9cm,height=4.5cm]{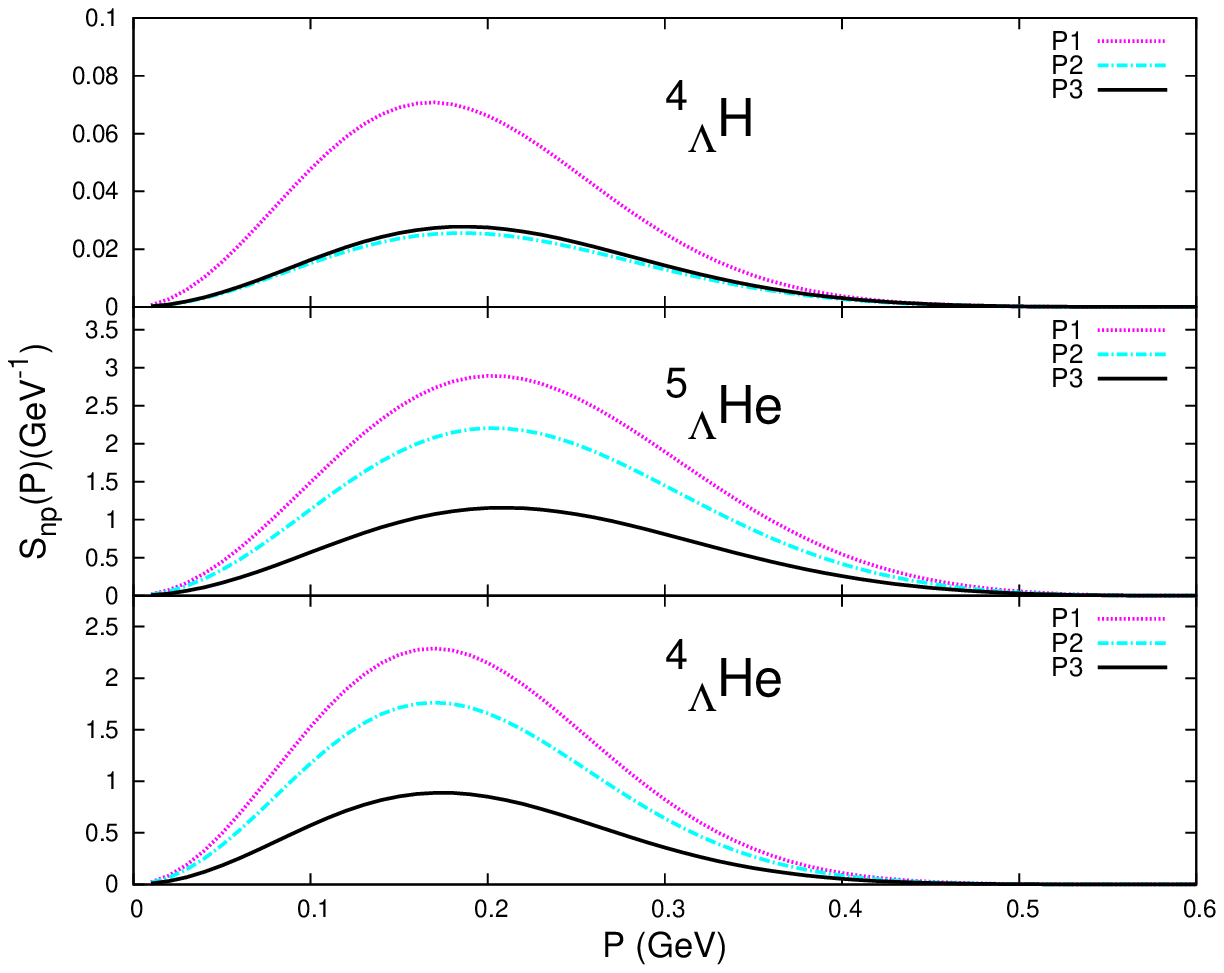}&
\hspace{2cm}\includegraphics[width=4.9cm,height=4.5cm]{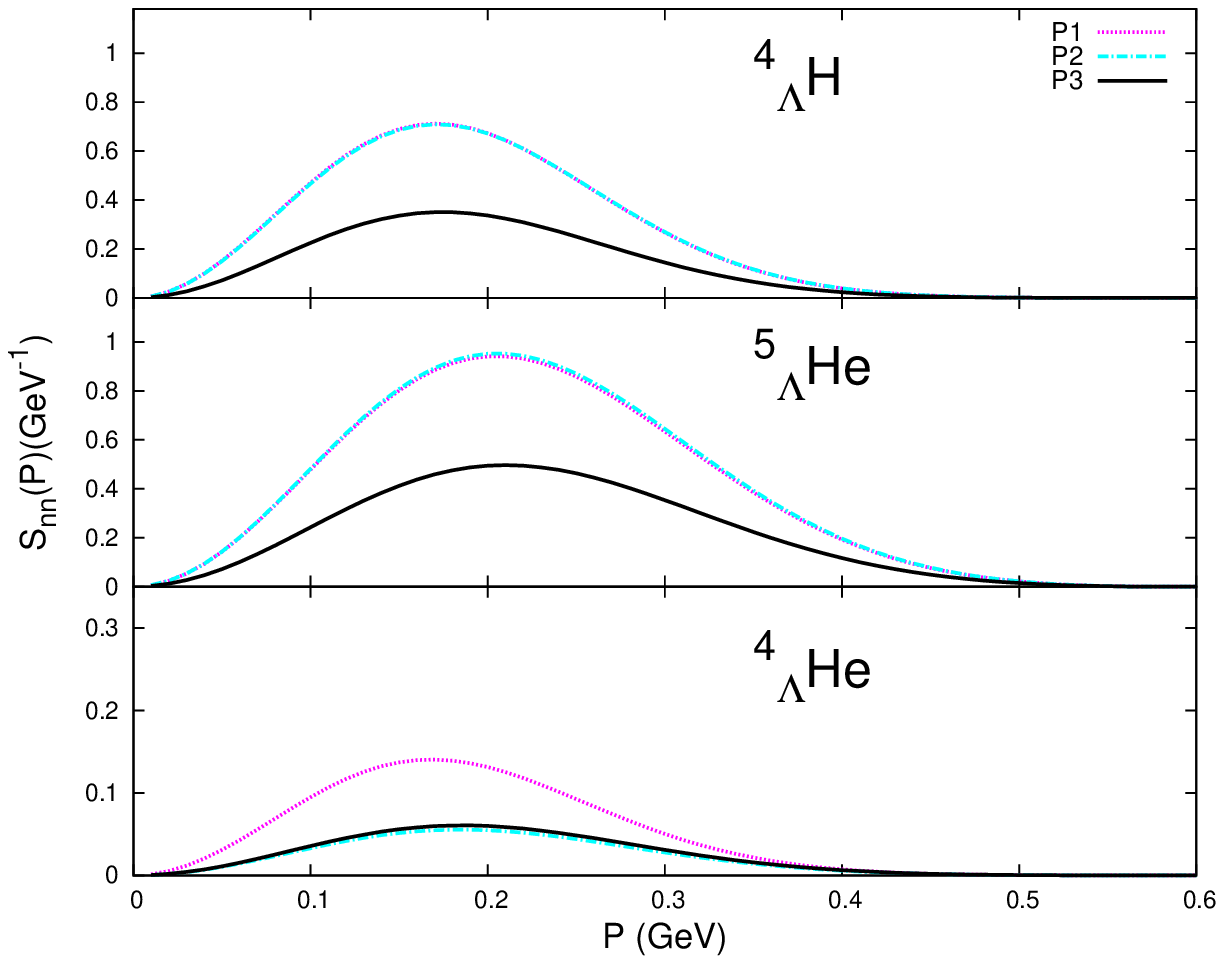}
\end{tabular}
\vspace{0.51cm}
 \caption{\label{F4}(Color online)
Calculations  of   the c.m. momentum
correlations  $S_{nN}(P)$  of $np$ (left panel) and $nn$ (right panel) pairs  for three different
parametrizations.}
\end{figure}

The results for the coincidence spectra as a function of the c.m. momentum $P$ are shown in the Fig. \ref{F4}.
Theoretically, they behave as
 \br &&S_{nN}(P)\sim
P^2\sqrt{P_N^2-P^2}e^{-(Pb)^2/2},
\label{8}\er
from $P=0$ up to the    maximum values of c.m. momenta
\be
{P}_N=2\sqrt{\frac{A-2}{A}{\rm M}\Delta_N}.
 \label{9}\ee
The maximum of the correlation spectra $S_{nN}(P)$ occurs at the value of $P$ equal to
\br
P_N^\uparrow&=&\left[\frac{b^2P_N^2+3-\sqrt{{b^4P_N^4-2b^2P_N^2+9}}}{2b^2}\right]^{1/2}.
\label{10}\er

For example, the  \he~ spectrum $S_{np}(P)$ goes up to  the maximum value of the c.m. momentum
 $P_p=0.546$ GeV, while its maximum occurs  at $P_p^\uparrow=0.165$ GeV.
 It is unfortunate that in the literature there are
 no data on the spectra $S_{nN}(P)$  for any of the three hypernuclei
 in order to make comparison with theoretical predictions.

\begin{table}[htpb]
\caption{Comparison between the experimental  transition rates,
derived from Figs. \ref{F1}, \ref{F2}, and  \ref{F3} by means  of Eq. \rf{3},
 and the IPSM calculation, for  the   OME potential P3.} \label{T1}
\begin{center}
\begin{tabular}{|c|c|c|c|}
\hline
&$\Gamma_{p}$&$\Gamma_{n}$&$\Gamma_{p}+\Gamma_{n}$\\
\hline
\he, BNL~\cite{Pa07} &&&\\
Fig.\ref{F1}&$0.185\pm 0.011$&$-0.029\pm0.021$&$0.156\pm0.032$\\
Fig.\ref{F2}&$0.242\pm 0.101$&$0.084\pm0.032$&$0.326\pm0.133$\\
Fig.\ref{F3}&$0.176\pm 0.021$&$0.074\pm0.019$&$0.250\pm0.040$\\
Theory P3&$0.179$&$0.012$&$0.191$\\
bare value~\cite{Pa07}&$0.189\pm 0.028$&$\le 0.035$&$0.177\pm0.029$\\
\hline
 \He, KEK~\cite{Ok04,Ka06,Ki06,Bh07} &&&\\
Fig.\ref{F1}&$0.161\pm 0.008$&$0.219\pm0.016$&$0.380\pm0.024$\\
Fig.\ref{F2}&$0.129\pm 0.062$&$0.118\pm0.038$&$0.247\pm0.106$\\
Theory P3&$0.281$&$0.121$&$0.402$\\
bare value~\cite{Ki06,Bh07}&$0.296\pm0.020$&$0.133\pm0.020$&$0.429\pm0.017$\\
 \hline
   \end{tabular}
\end{center}
\end{table}
The identity \rf{7} and similar relationships for protons  allow us to relate
the experimental transition probabilities derived from different spectra through the Eq.  \rf{3}
 and those corresponding to other observables. They are compared in Table \ref{T1},
and although  labelled as $\Gamma_{p}$, and $\Gamma_{n}$ they include
not only contributions arising from $\Gamma_1$ but also from $\Gamma_2$, as well as
the effects of the FSIs.
As can be see the results coming from different spectra differ considerably, and the answer to these discrepancies
must come from the experimental side.
For the sake of completeness in  the same table we also show the values  of the "bared transitions rates",
which were extracted from the experimental data~\cite{Pa07,Ki06,Bh07} excluding the effects of both
the $2N$-NM decay, and the FSIs.

\section{Final Remarks}
Single and coincidence spectra of the NM weak decay of light
hypernuclei have been evaluated in a systematic way for the first
time. We have considered the 1N induced processes only, omitting
entirely the 2N induced  events, as well as the effects
of the FSIs. However, we have discussed in detail the recoil effect,
showing that it is very important in the description of the spectra of light hypernuclei.
 For the theoretical framework we have used the IPSM
with three different parametrizations for the transition
potential. The comparison with data  suggests that the
soft $\pi+K$ exchange model, introduced in Ref.~\cite{Ba09},  could be a good
starting point to describe the dynamics of the NMWD in
\he, and \He~hypernuclei.  Such a statement is supported by the
results shown in Fig. \ref{F1} for the kinetic energy spectra, and in Figs. \ref{F2}, and  \ref{F3} for $pn$
coincidence spectra.
In spite of this agreement we feel that  a more realistic description of the SRC, such as done in Refs.
\cite{In97,Ok98,Sa00,Sa02,Ok05,Sa05} by means of the DQ weak transition potential,
may help us to better understand the baryon-baryon
strangeness-flipping  interaction.

Definitively, the coincidence $nn$ data
 can not be explained theoretically neihter in \he~ nor in \He. On the other hand, due to
the lack of  data nothing can be said regarding $^4_\Lambda$H. We note,
however, that both the single and coincidence neutron spectra in
this case are of the same order of magnitude as in the other two
hypernuclei, and that therefore it would be extremely  useful to
measure them experimentally.

Despite having achieved a relatively fair agreement with the data for several exclusive
observable, such as proton and neutron kinetic energy spectra and the $pn$ coincidence
spectra, the question arises whether
it is legitimate to use the SM, and
in particular the harmonic oscillator wave functions for systems as small as are $^4_\Lambda$H, \he, and \He.
Moreover, as pointed out by J-H. Jun~\cite{Ju01}, the residual nucleons are in a unbound state
for some channels, and therefore they are very different from the initial nucleons  which
do not take part  in the decay process. Then,
it could  be appropriate  to treat  the four- and five-body nature of these
NMWD explicitly, as was done for instance in the evaluation of separation energies
in single and double strange hypernuclei~\cite{Ku02,Fi02,No02,Ne02,Ne05,Hi10},
and in {\it ab initio} calculations
of four-nucleon scattering~\cite{Del12}.
(The shell model yields quite nice results for light $S=-1,-2$ hypernuclei~\cite{Ka03,Ga11}.)
It would be very interesting to analyze whether it  is possible to account
 for the experimental nn coincidence spectra  through such microscopic calculations.
An undertaking of this nature is, however, beyond our present means.

\begin{center}
{\bf ACKNOWLEDGEMENTS}
\end{center}
 This work was partly supported by  Argentinean agencies CONICET (PIP 0377) and FONCYT (PICT-2010-2680),
as well as by the Brazilian agency FAPESP (CONTRACT 2013/01790-5).
I thank Joe Parker for very useful discussions.

\end{document}